\documentclass[aps,prl,reprint,superscriptaddress]{revtex4-2}
\usepackage{amsfonts}
\usepackage{amsmath}
\usepackage{amssymb}
\usepackage{graphicx}
\usepackage{epstopdf}
\usepackage{color}
\usepackage{bbold}
\usepackage[colorlinks, citecolor=red]{hyperref}

\renewcommand{\vec}[1]{\boldsymbol{#1}}

\begin{document}

\title{Quantum spin decoherence theory of magnetoresistance in mesoscopic  ferromagnets and its applications}

\author{Xian-Peng Zhang}

\affiliation{Centre for Quantum Physics, Key Laboratory of Advanced Optoelectronic Quantum Architecture and Measurement (MOE), School of Physics, Beijing Institute of Technology, Beijing, 100081, China}
\affiliation{International Center for Quantum Materials, Beijing Institute of Technology, Zhuhai, 519000, China}

\affiliation{Department of Physics, Hong Kong University of Science and Technology, Clear Water Bay, Hong Kong, China}

\author{Xiangrong Wang}

\affiliation{Department of Physics, Hong Kong University of Science and Technology, Clear Water Bay, Hong Kong, China}

\author{Yugui Yao}
\email{ygyao@bit.edu.cn}
\affiliation{Centre for Quantum Physics, Key Laboratory of Advanced Optoelectronic Quantum Architecture and Measurement (MOE), School of Physics, Beijing Institute of Technology, Beijing, 100081, China}

\affiliation{International Center for Quantum Materials, Beijing Institute of Technology, Zhuhai, 519000, China}

\begin{abstract}
Quantum decoherence is the key mechanism determining whether quantum effects can manifest in quantum computation and transport, and mastering decoherence is central to designing and operating functional quantum devices. Here, we present a quantum spin decoherence theory of magnetoresistance (MR) from an open-quantum system perspective. Importantly, even when the spin-up and -down species have the same density of states,  magnon MR, anisotropic MR, and Hanle MR  still emerge in \textit{mesoscopic}  ferromagnets, which arise from the magnon-induced spin flip, spin relaxation anisotropy, and Hanle spin precession of itinerant electrons, respectively. The theory not only predicts the magnetic field and temperature dependencies of MR, which are related to spin relaxation 
time and spin-exchange field, but also obtains the universal cosine-square law of anisotropic MR. Moreover, we reveal diverse behaviors of the MR effects that enable the simple detection of the spin-exchange coupling strength via an electrical measurement. Our theory advances the understanding of the fundamental physics of MR in mesoscopic ferromagnets, revealing how it enables electrical probing of quantum decoherence—the decisive factor in nanoscience and nanotechnology.
\end{abstract}

\maketitle

\textit{Introduction--}Magnetoresistance (MR), one of the most fundamental yet still mysterious phenomena in condensed matter physics since discovered in 1856 by Kelvin~\cite{thomson1857xix}, is highly significant and technologically crucial for the modern microelectronics industry~\cite{dieny2020opportunities}. The origin of this century-old effect is known as the spin-exchange coupling (SEC) between local moments and itinerant electrons in magnetic materials~\cite{mott1936electrical,mott1936resistance,mott1964electrons,goodings1963electrical,yosida1957anomalous,raquet2002electron,mihai2008electron,zhang2023extrinsic,zhang2022microscopic}, while microscopic mechanisms have not been fully understood due to its intrinsic complexity as a many-body and open-quantum problem~\cite{zhang2025open}. The previous theories of MR primarily focused on the momentum relaxation of itinerant electrons~\cite{ritzinger2023anisotropic}. For example, the 
anisotropic MR, i.e., the change of resistance with the orientation of the magnetization relative to the electric current~\cite{mcguire1975anisotropic,campbell1970the,ebert1996anisotropic,smit1951magnetoresistance,van1959anisotropy,trushin2009anisotropic,wang2023theory,wang2024theory}, originates from anisotropic \textit{momentum relaxation} arising from either the magnetization-dependent scattering rate of itinerant electrons~\cite{mcguire1975anisotropic,campbell1970the,ebert1996anisotropic,smit1951magnetoresistance,van1959anisotropy,trushin2009anisotropic} or the magnon population of the local moments~\cite{goodings1963electrical,yosida1957anomalous,raquet2002electron,mihai2008electron} (i.e., magnon MR~\cite{khvalkovskiy2009high,nguyen2011detection}). Here we identify a distinct and ubiquitous mechanism of MR—SEC–induced \textit{spin decoherence}—which constitutes a previously unrecognized contribution in mesoscopic ferromagnets. The associated MR effect enables electrical measurement of quantum decoherence, whose control is essential for fundamental studies and emerging applications in nanoscience and nanotechnology~\footnote{In the post-Moore era, nanoscience and nanotechnology are crucial for exploiting quantum, spin, and interfacial phenomena at the nanoscale, enabling novel, energy-efficient device paradigms beyond conventional transistor scaling.}.

Quantum coherence, at the heart of quantum mechanics, emerges when an electron exists in multiple quantum states simultaneously~\cite{zurek2003decoherence,streltsov2017colloquium}, such as spin superposition state $\vert\psi_{\vec{p}}(t)\rangle =\sum_{s=\uparrow,\downarrow}c_{s\vec{p}} (t)|s\vec{p}\rangle$ 
with $c_{s\vec{p}} (t)=c_{s\vec{p}} e^{-i\epsilon_{s\vec{p}}t/\hbar}$. 
Density matrix $\varrho_{\vec{p}}(t)=\vert\psi_{\vec{p}}(t)\rangle \langle\psi_{\vec{p}}(t)\vert$ offers a natural framework:  diagonal elements $\varrho^{ss}_{\vec{p}}(t)=\vert c_{s\vec{p}}(t)\vert^2$ encode population, while off-diagonal ones $\varrho^{\uparrow\downarrow}_{\vec{p}}(t)=c^{}_{\uparrow\vec{p}}(t)c^{*}_{\downarrow\vec{p}}(t)$ quantify coherence. This coherence enables interference to generate novel quantum transport~\cite{chang2023colloquium,sinova2015spin}. However, spin coherence decays due to interactions with local moments~\cite{zhang2025open,zhang2019theory,gomez2020strong,oyanagi2021paramagnetic,zhang2024microscopic,zhang2022extrinsic} through two processes~\cite{zhang2025anomalous,zhang2025theory,breuer2002theory,nielsen2010quantum,abragam1961principles}: i) spin dephasing with timescale $\tau_\phi$ randomizes the relative phase $\left\langle e^{\text{ang}[c_{\uparrow\vec{p}}(t)/c_{\downarrow\vec{p}}(t)]i} \right\rangle \sim e^{-t/\tau_\phi}$~\footnote{where $\langle \cdots \rangle$ denotes ensemble averaging.}; ii) population relaxation with timescale $\tau_{\Vert}$ reduces the amplitudes of the superposition state $\langle |c_{s\vec{p}}(t)c^{*}_{s\vec{p}}(t)| \rangle \sim e^{-t/\tau_{\Vert}}$, simultaneously resulting in a decay of the coherence magnitude   
$\langle |c_{\uparrow\vec{p}}(t)c^{*}_{\downarrow\vec{p}}(t)| \rangle \sim e^{-t/(2\tau_{\Vert})}$~\cite{zhang2019theory}. The spin decoherence time $\tau_{\perp}$, governing the decay of the off-diagonal density matrix element, $\langle c_{\uparrow\vec{p}}(t)c^{*}_{\downarrow\vec{p}}(t) \rangle \sim e^{-t/\tau_{\perp}}$, combines both processes: $1/\tau_{\perp}=1/(2\tau_{\Vert})+1/\tau_{\phi}$~\cite{breuer2002theory,nielsen2010quantum,abragam1961principles}. Notably, spin decoherence  exhibits strong magnetic-field ($B$) and temperature ($T$) dependence~\cite{gomez2020strong,oyanagi2021paramagnetic}, yet is absent from existing theories of MR in \textit{mesoscopic} ferromagnets.

\begin{figure}[t!]
\begin{center}
\includegraphics[width=0.48\textwidth]{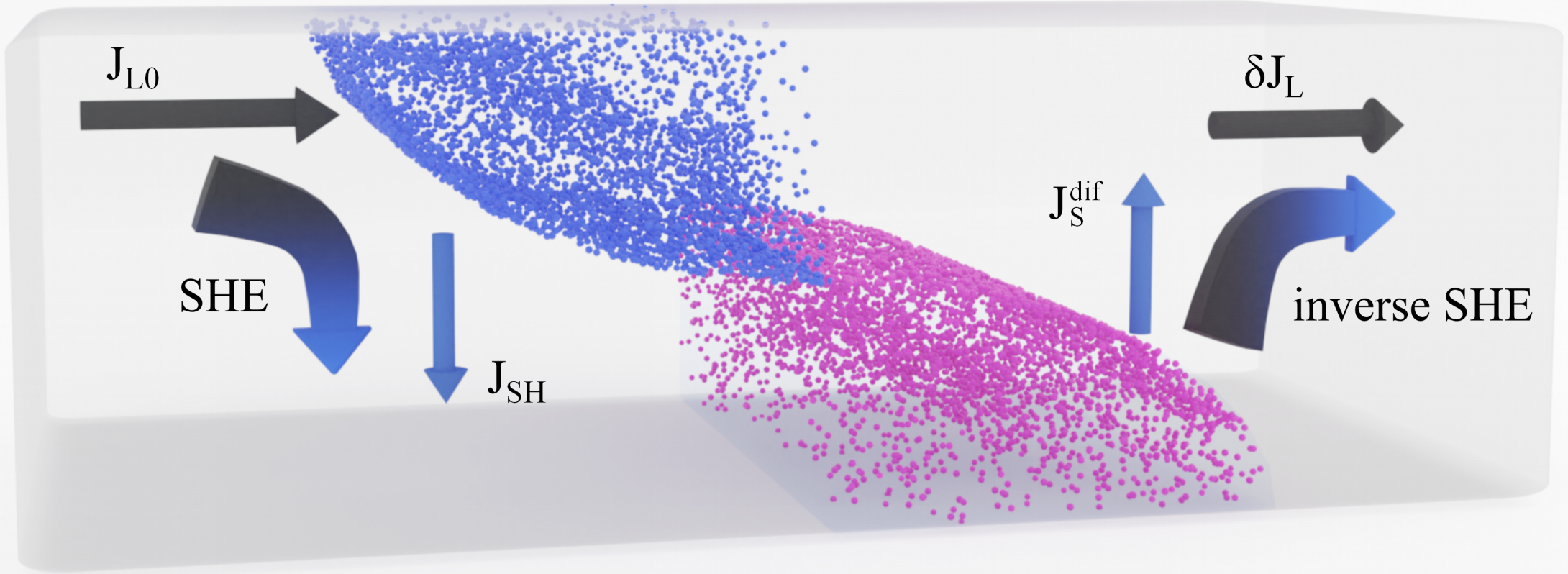} 
\end{center}
\caption{(Color online) The MR effect arises from a two-step charge–spin conversion process in a ferromagnet monolayer. }
\label{STORY}
\end{figure}

Here, we develop a quantum spin decoherence theory of MR from an open-quantum system perspective that accounts for the $B$ and $T$ dependencies of the spin decoherence processes. We derive a universal cosine-square formula of MR that elucidates the microscopic mechanisms of magnon, anisotropic, and Hanle MR arising from the magnon-induced spin flip, anisotropic spin relaxation, and spin precession of itinerant electrons. The related MR enables the simple detection of the microscopic SEC strength and the resulting  spin decoherence.

\textit{Model--}The MR effect, at its core, deals with the interplay between local moments and itinerant electrons. We  consider a ferromagnetic material described by the many-body Hamiltonian  
$H=H_{e}+H_{m}+V_{em}$. The itinerant electrons' Hamiltonian  is given by $H_e=p_n^2/2m+\hbar\omega_B {\bf m }\cdot\vec{\mathbf{\sigma}}_n-(\hbar/4m^2c^2)\vec{\sigma}_n \cdot [\vec{p}_n \times \vec{\nabla} V_{\text{so}}(\vec{r}_n)]$ with  
$\omega _{B}=\textsl{g}\mu _{B}B/\hbar$ and $\mathbf{m}=\vec{B}/B$, where $\textsl{g}\simeq 2$ is the $g$-factor, $\mu _{B}$ is the Bohr magneton, $\hbar$ is the reduced Planck constant, $\vec{\sigma}_n=(\sigma_n^x,\sigma_n^y,\sigma_n^z)$ is the vector of Pauli matrices representing the $n$th itinerant electron with position $\vec{r}_n$ and momentum $\vec{p}_n$, and $V_{\text{so}}(\vec{r}_n)$ includes both intrinsic and extrinsic spin–orbit potential~\cite{winkler2003spin}. Hereafter,  repeated indices are summed over. We also include a term describing the coupling of  the local moments themselves and with external magnetic field $H_{m}=\textsl{g}\mu _{B}\mathbf{%
S}_{j}\cdot \vec{B}-J_{i j} \mathbf{S}_{i} \cdot \mathbf{S}_{j}$,
where $J_{ij}>0$ is the coupling constant of the Heisenberg ferromagnet.  
Finally, $V_{em}$ describes the effect of the local moments with scattering potential~\cite{mott1936electrical,mott1936resistance,mott1964electrons,goodings1963electrical,yosida1957anomalous,raquet2002electron,mihai2008electron,zhang2023extrinsic,zhang2022microscopic}
\begin{equation} \label{spin-exchange}
 V_{em}=-\mathcal{J}_{sd} \vec{\sigma}_n\cdot\mathbf{S}_j\delta(\vec{r}_n-\vec{R_j}) ,
\end{equation}
where $\mathbf{S}_j$ is the spin of the $j$th local moment with position $\vec{R_j}$ and 
positive (negative) $\mathcal{J}_{sd}$ corresponds to ferromagnetic (antiferromagnetic) SEC.

\textit{Quantum spin decoherence theory--}The key role of the SEC, in the MR effect, lies in its induction of anisotropic spin relaxation for free electrons~\cite{zhang2025open,zhang2019theory,zhang2022extrinsic,zhang2024microscopic,gomez2020strong,oyanagi2021paramagnetic}. During propagation inside the mesoscopic ferromagnet, the spin exchange is inherently accompanied by the spin decoherence~\cite{zhang2019theory,zhang2022extrinsic,zhang2024microscopic}.  Based on the theory of open-quantum system~\cite{breuer2002theory,zhang2025open}, we analytically derive the spin decoherence from the Liouville-von Neumann equation via time-convolutionless projection operator method, where the quantum nature of local moments, treated as quantum bath, is significant for the $B$ and $T$ dependencies of the itinerant electrons' spin decoherence. Note that spin decoherence has two distinct pathways (i.e., dephasing and  relaxation). The spin dephasing processes randomize the relative phase with spin dephasing time~\cite{zhang2025open}
\begin{align}  \label{dephasingSRT}
	\frac{1}{\tau_{\phi}} &= \frac{1}{2\tau_0} 
	+\frac{\pi}{\hbar}   n_{\text{S}}  \nu_F \mathcal{J}^2_{sd}\langle S^2_{\Vert
}\rangle,
\end{align}
where $\nu_F$ is the density of states per spin at the Fermi level assumed to be the \textit{same} for spin-up and -down species~\footnote{\textit{\textquotedblleft The density of states is different for majority and minority
spins"} is the necessary condition for the previous MR theory based on momentum relaxation time that strongly relies on the density of state. In our theory, focusing on the anisotropic spin relaxation time, the density of state for spin-up and spin-down species can be the same. Notably, our spin decoherence theory still works for different case.}, $n_{\text{S}}$ is the density of local moments, and $S_{\Vert }$ is the spin component in the direction of the magnetization, which is assumed to be collinear to $\vec{B}$~\footnote{This is a widely used approximation for simplicity when we deal with the MR irrelevant to hysteresis and magnetostatic and magneto-crystalline anisotropy. Of course, following our method, we can directly include these effects by considering non-collinear case.}.
The first term, $\tau_0^{-1}$, denotes the isotropic spin relaxation rate unrelated to local moments~\footnote{It originates from the spin-orbit potential $V_{\text{so}}$ and is assumed to be isotropic and independent of temperature ($T$).}. The second term accounts for the spin dephasing rate arising from the SEC-induced scattering processes, during which the transverse spin component vanishes due to phase mismatch. On the other hand, the coherence also depends on the amplitude ($c_{\sigma\vec{p}}$) of the superposition state, whose decay is characterized by the longitudinal spin relaxation time~\cite{zhang2025open} 
\begin{align} \label{LongiSRT}
\frac{1}{\tau_{\Vert}}&= \frac{1}{\tau^{}_0}
+\frac{\pi}{\hbar}   n_{\text{S}}  \nu_F \mathcal{J}^2_{sd} \beta\epsilon_L
 n_{B}(\epsilon_L)[S(S+1)-\langle S^2_{\Vert
}\rangle-\langle S_{\Vert
}\rangle].
\end{align}
Here, $n_B(\epsilon^{}_L) = 1/(e^{\beta\epsilon^{}_L} - 1)$ is the Bose-Einstein distribution function with inverse temperature $\beta = 1/(k_BT)$ and  the effective Zeeman energy $\epsilon_{L}=\hbar\omega_B-\langle S_{\Vert}\rangle\sum_{i} J_{ij} $ of the local moments. 
The longitudinal spin relaxation arises from the SEC-induced spin-flip processes, thereby related to the magnon emission/absorption~\cite{zhang2024microscopic}. 
The coherence amplitude, as a product of two amplitudes of the superposition state, decays with spin decoherence time~\cite{zhang2025open} 
\begin{align} \label{TransSRT}
\frac{1}{\tau_{\perp}}&=\frac{1}{2\tau_0}+\frac{1}{2\tau_{\Vert}} 
	+\frac{\pi}{\hbar}   n_{\text{S}}  \nu_F \mathcal{J}^2_{sd}\langle S^2_{\Vert
}\rangle,
\end{align}
consistent with the standard relation $1/\tau_{\perp}=1/(2\tau_{\Vert})+1/\tau_{\phi}$ in quantum-computation and condensed-matter decoherence theory~\cite{breuer2002theory,nielsen2010quantum,abragam1961principles}. Coherent quantum transport requires minimizing decoherence. Controlling it is key to understanding quantum behavior and building practical quantum devices. Notably, the spin decoherence~\eqref{TransSRT} [including dephasing \eqref{dephasingSRT} and relaxation~\eqref{LongiSRT}] strongly depends on $B$ and $T$ via $\langle \text{S}_{\Vert} \rangle$ and $\langle \text{S}^2_{\parallel} \rangle$, and is responsible for the $B$- and $T$-dependent MR effect. Next, we show how spin decoherence participates the MR in mesoscopic ferromagnets that allows for measuring decoherence.

Mesoscopic transport, with broken translational symmetry, itself is a challenging problem in analytical physics: the system thickness $d_N$ is comparable to the spin-diffusion length, and interfacial effects such as edge spin accumulations become decisive for quantum transport. In our case, MR originates from the interplay of the spin Hall effect (SHE) and its inverse effects caused by the skew scattering of the extrinsic spin–orbit coupling~\cite{maekawa2017spin,sinova2015spin}, as illustrated in Fig.~\ref{STORY}. First, a longitudinal charge current $\vec{J}_{\mathrm{L0}}=\sigma_{\mathrm{D}}\vec{E}$ is converted into a transverse drift spin current $\vec{J}_{\mathrm{SH}}=\theta_{\mathrm{SH}}(\hat{y}\times \vec{J}_{\mathrm{L0}})$ via the SHE  (left-curved arrow). Here, $\theta_{\mathrm{SH}}\propto V_{\text{so}}$~\cite{sinova2015spin,zhang2024altermagnet} denotes the spin Hall angle, $\sigma_{\mathrm{D}}$ is the Drude conductivity, and $\vec{E}$ is the applied electric field along $\hat{x}$. Assuming system is uniform in $x-y$ plane, the resulting spin current, polarized along $\hat{y}$ and flowing along $\hat{z}$, produces a spin accumulation $\mu_{s}^{y}(z)$ that drives a counterflowing diffusive spin current, $J_{\mathrm{S}}^{\mathrm{dif}}=-\frac{\sigma_{\mathrm{D}}}{2e}\partial_{z}\mu_{s}^{y}(z)$, with $e$ being the charge of electron. Second, this diffusive spin current is converted back into charge current in the same direction as $\vec{J}_{\mathrm{L0}}$ through the inverse SHE  (right-curved arrow in Fig.~\ref{STORY}), which yields a longitudinal resistivity correction $
\Delta\rho_{\mathrm{L}}=-2\theta_{\mathrm{SH}}\sigma_{\mathrm{D}}E-\theta_{\mathrm{SH}}\sigma_{\mathrm{D}}[\mu_{s}^{y}(d_N)-\mu_{s}^{y}(0)]/d_N
$~\cite{dyakonov2007magnetoresistance,zhang2019theory}, averaged over thickness direction. Consequently, the edge spin-$y$ accumulation  governs the MR of mesoscopic ferromagnets.

Following the derivations of Ref.~\cite{zhang2025open}, we obtain the microscopic expression of the widespread phenomenological formula of the longitudinal resistivity
\begin{equation}  \label{SL}
\rho _{L}\simeq \rho _{\mathrm{L0}}+ 2\theta _{\mathrm{SH}}^{2}\rho _{\mathrm{L0}}+\theta _{\mathrm{SH}}^{2}\Delta
\rho _{0}+\theta _{\mathrm{SH}}^{2}\Delta \rho _{1}\left( 1-\hat{m}%
_{y}^{2}\right) ,
\end{equation}%
with
\begin{equation} \label{vvh0}
\frac{\Delta \rho _{0}}{\rho _{\mathrm{L0}}}=-\frac{2\ell _{\Vert }}{d_{N}}\tanh \left( \frac{%
d_{N}}{2\ell _{\Vert }}\right),
\end{equation}%
\begin{equation} \label{vvh1}
\frac{\Delta \rho _{1}}{\rho _{\mathrm{L0}}}=\frac{2\ell _{\Vert  }}{d_{N}}\tanh \left( \frac{%
d_{N}}{2\ell _{\Vert  }}\right)-\mathrm{Re}%
\left[ \frac{2\Lambda}{d_{N}}\tanh \left( \frac{%
d_{N}}{2\Lambda}\right)\right],
\end{equation}%
where $\rho _{\mathrm{L0}}=1/\sigma_{\text{D}}$. The SEC-induced longitudinal and transverse spin diffusion lengths read $%
\ell _{\mathrm{\parallel }}=\sqrt{\mathcal{D}\tau _{\mathrm{\parallel }}}$ and $\ell _{\mathrm{\bot }}=\sqrt{\mathcal{D}\tau _{\mathrm{\bot }}}$, respectively, where $\mathcal{D} $ is diffusion coefficient. Besides, the SEC shifts the Hanle spin precession frequency $\omega
_{L}=\omega_{B}-\textsl{g}\mu_B\mathcal{B}_{sd}/\hbar$ by spin-exchange field $\mathcal{B}_{sd}=n_{
\mathrm{S}}\mathcal{J}_{sd}\langle S_{\Vert }\rangle/( \textsl{g}\mu_B)$ and spin precession length is given by $\ell _{\mathrm{L}}=\sqrt{\mathcal{D}/\omega _{L}}$. The spin-mixing length $
\Lambda^{-2}=\ell _{\mathrm{\bot }}^{-2}+i\ell _{\mathrm{L}}^{-2}$ depends on spin decoherence and precession. Equations (\ref{SL}-\ref{vvh1}), as the central result, not only microscopically explain the universal cosine-square law of anisotropic MR with the magnetization direction ($\hat{m}_y\equiv\sin\alpha$)  but also quantitatively describe $B$- and $T$-dependent MR.  Importantly, both $T $ and $B$ dependencies of $\Delta
\rho _{0}$ and $\Delta
\rho _{1}$, i.e., Eqs.~\eqref{vvh0} and \eqref{vvh1}, are
reflected in spin precession length ($\ell _{\mathrm{L}}$), spin diffusion lengths ($\ell _{\Vert,\perp} $) through the spin-exchange field and spin relaxation time, respectively. The dependence of resistance on $T$ arises from $\langle S_{\Vert}\rangle $ and $\langle S_{\Vert}^{2}\rangle $, while that of $B$ has an extra
channel - the spin precession frequency $
\omega _{B}(\propto B)$.

\begin{figure}[t]
\begin{center}
\includegraphics[width=0.48\textwidth]{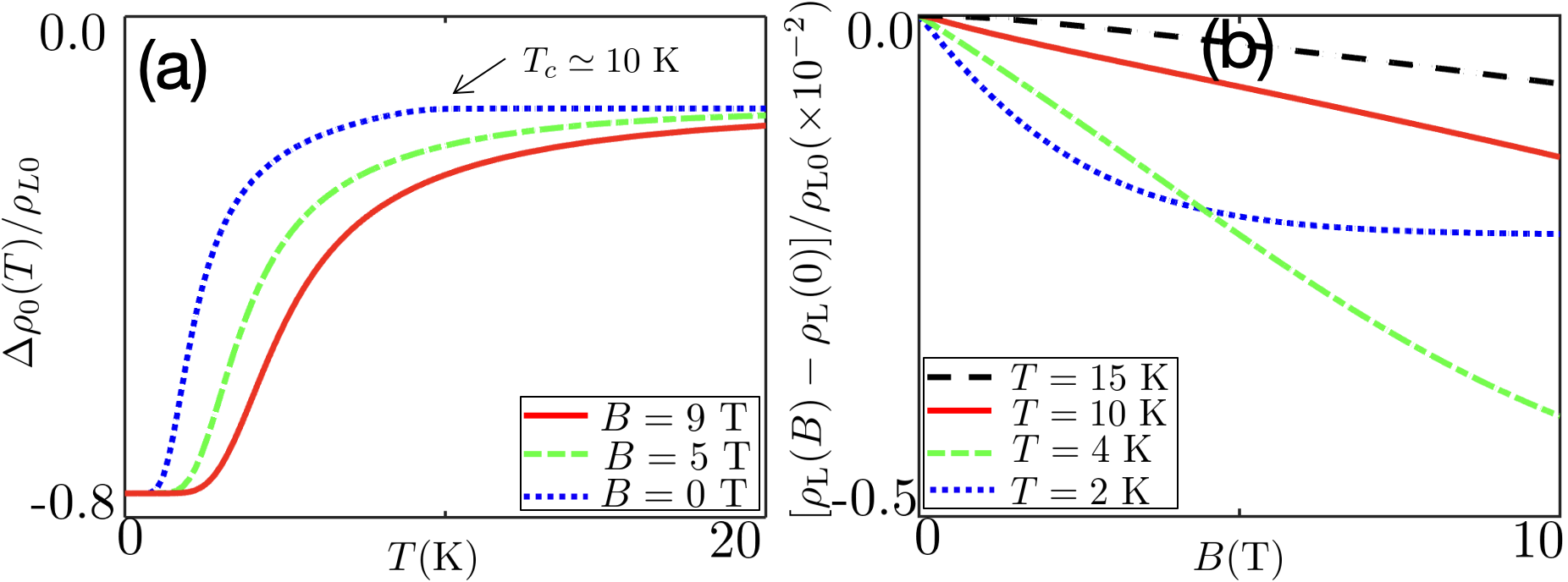} 
\end{center}
\caption{(Color online) Magnon MR at $\hat{m}_y=1$. (a,b) $
\Delta\rho _{0}$ vs (a) temperature $T$  and (b) magnetic field $B$. We set  $n_{\mathrm{S}}\mathcal{J}_{sd}=8$ meV and a $T_c=10$ K. Other parameters: $\theta _{%
\mathrm{SH}}=0.1$, $S=2$, $\ell _{0}=3.0$ nm, $d_{N}=5$ nm, $E_F=1.0 $ eV, $m_{F}=1.0$ $m^0_e$, $\rho_{L0}=2.0\times 10^{6}$ $\Omega\cdot m$, and $\mathcal{D}=1.0*10^{-6}$ m$^2$/s.}
\label{MMR}
\end{figure}

Then, we investigate the microscopic mechanisms of various MR effects using our formulas (\ref{SL}-\ref{vvh1}). Note that the SEC  in magnetic materials is ubiquitous and profoundly affects the MR effects. At $\mathcal{J}_{sd}=0$, the spin diffusion lengths become isotropic (i.e.,  $\ell_{\Vert,\perp}=\ell_0=\sqrt{\mathcal{D}\tau_0}$) and we recover the previous theory of  Hanle MR~\cite{velez2016hanle} where $\Delta \rho_1\propto B^2$ for small $B$. Next, we consider $\mathcal{J}_{sd} \neq 0$. i) The SEC shifts the Hanle spin precession frequency $\omega_L=g\mu_B(B-\mathcal{B}_{sd})$, introduces a finite value of $\Delta \rho _{1}\propto (B-\mathcal{B}_{sd})^2$ even when the anisotropic spin relaxation is artificially removed by setting $\ell_{\Vert}= \ell_{\perp}$, and contributes to a shifted Hanle MR. Thus, our theory, different from the previous Hanle MR independent of $T$,  effectively includes $T$ and $B$ dependencies of MR through $\mathcal{B}_{sd}$. ii) The SEC causes anisotropic spin diffusion lengths ($\ell_{\Vert}\neq \ell_{\perp}$), produces a finite value of $\Delta \rho _{1}=\frac{2\ell _{\Vert  }}{d_{N}}\tanh \left( \frac{%
d_{N}}{2\ell _{\Vert  }}\right)-\frac{2\ell _{\perp  }}{d_{N}}\tanh \left( \frac{%
d_{N}}{2\ell _{\perp }}\right)$ when artificially setting $\omega_L=0$, and accounts for anisotropic MR. Thus, our microscopic theory, exceeding the previous phenomenological theory of anisotropic MR, adequately captures the $T$ and $B$ dependencies of $\Delta\rho_{1}$ through the spin relaxation time [Eqs.~\eqref{LongiSRT} and \eqref{TransSRT}]. iii) The $\hat{m}$-independent resistance, quantified by $\Delta\rho _{0}$, depends only on the longitudinal spin relaxation time~\eqref{LongiSRT} originating 
from spin-flip processes, and this MR is unambiguously associated with magnon emission and absorption, thus referring to  magnon MR. Overall, we demonstrate the magnon, anisotropic, and Hanle MR originating from the magnon-induced spin flip, anisotropic spin relaxation, and Hanle spin precession of itinerant electrons, respectively.

\emph{Magnon MR}-For magnetization parallel to the spin polarization of the SHE  (i.e., $\hat{m}_y=\pm 1$), only magnon MR survives, i.e., $\rho _{\mathrm{L}}\simeq \rho _{\text{L0}}+2\theta _{\mathrm{SH}}^{2}\rho _{\mathrm{L0}}+\theta _{\mathrm{SH}}^{2}\Delta
\rho _{0}$. Figure~\ref{MMR}(a) [\ref{MMR}(b)] plots the $T$ [$B$] dependencies of $\Delta\rho _{0}$ [$\rho _{\mathrm{L}}$]. To qualitatively illustrate the magnon MR, we consider the case of small enough $B$ [blue curve in Fig.~\ref{MMR}(a)]. For $T>T_c$, the local-moment spins are randomly oriented, and the majority of edge spin-$y$ accumulations are dissipated via spin flips mediated by magnon emission and absorption, where $T_c$ is the Curie temperature. As a result, only a small fraction of the diffusive spin Hall current survives and is converted back into charge current, placing the system in a high-resistance state. In contrast, for $T\ll T_c$, the spins of the local moments are frozen along $\vec{B}$, suppressing spin flips and thus magnon emission/absorption, which drives the system into a low-resistance state.

\begin{figure}[t]
\begin{center}
\includegraphics[width=0.46\textwidth]{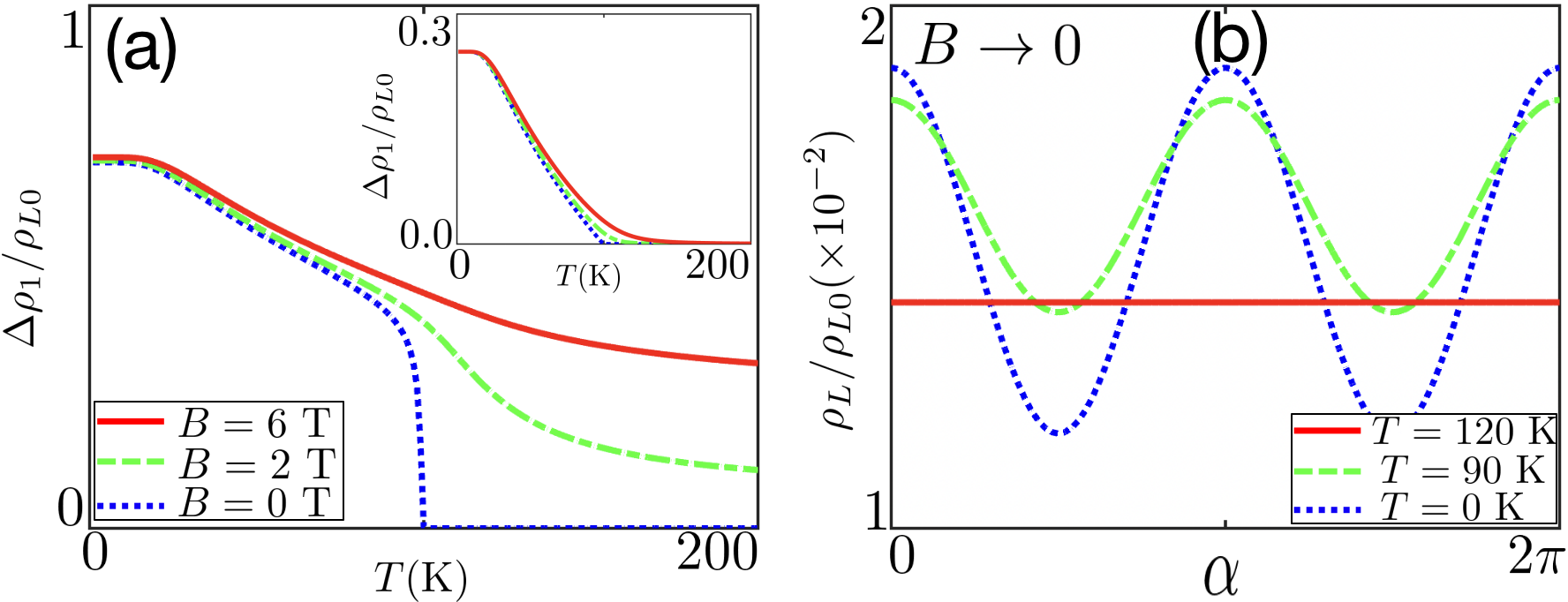} 
\end{center}
\caption{(Color online) Anisotropic MR. (a)   $
\Delta\rho _{1}$ vs $T$ for various $B$. The curve shape indicates two contributions: spin-exchange field and anisotropic spin relaxation. The spin-exchange field results in a sharp drop at the critical temperature when $B=0$ (blue curve). The contribution from anisotropic spin relaxation is represented in the inset, where $\mathcal{B}_{sd}$ is set to zero. (b) Resistivity  as a function of the magnetization direction ($\alpha$). We set $T_c=100$ K and $
n_{\mathrm{S}}\mathcal{J}_{sd}=1$ meV and other parameters are the same as for FIG.
\protect\ref{MMR}.}
\label{AMR}
\end{figure}

Quantitatively, $\Delta\rho _{0}$ initially exhibits a linear decrease following the expression $\Delta\rho _{0}=-2\ell_{\Vert}/d_N$ with increasing $\ell_{\Vert}/d_N$, and gradually approaches a saturation value of $-1$, in accordance with the asymptotic formula $\Delta\rho _{0}=-1+d^2_N/(12\ell^2_{\Vert})$ in the end. Note that $\ell_{\Vert}$ varies from $\sqrt{\mathcal{D}/(\tau^{-1}_{0}+   \Omega_{1})}$  to $\ell_0$  when cooling down the temperature from $T>T_c$  to $T\ll T_c$ at $B=0$ [blue curve in Fig.~\ref{MMR}(a)], where $\Omega_{1}=\frac{2\pi}{3\hbar}\nu_Fn_{\text{S}}\mathcal{J}^2_{sd}S(S+1)$. We obtain considerable modulation of magnon MR with $T$, i.e., $\vert \Delta\rho_{0}(T)-\Delta\rho_{0}(0)\vert/\rho_{\text{L0}}\sim 1$ [Fig.~\ref{MMR}(a)]. To increase the $B$ tunability of $\Delta\rho _{0}$ via the longitudinal spin relaxation time \eqref{LongiSRT}, we consider a small Curie temperature ($T_c=10$ K) and a large SEC ($
n_{\text{S}}\mathcal{J}_{sd}=8$ meV).  The latter leads to a large value of $\Omega_1$ and a small value of $\sqrt{\mathcal{D}/(\tau^{-1}_{0}+   \Omega_{1})}/d_N\simeq 0.07$.  Then, we obtain sizable $B$ modulation of magnon MR, i.e., $[\rho_{\text{L}}(B)-\rho_{\text{L}}(0)]/\rho_{\text{L0}}\sim 10^{-2}$, which always shows negative MR [Fig.~\ref{MMR}(b)] because increasing $B$ prevents the magnon-induced spin flip  and increases $\ell_{\Vert}$. The predicted characteristics qualitatively agrees with magnon MR experiments~\cite{nguyen2011detection}. Therefore, we attain a new mechanism of magnon MR owing to the interplay of the SHE  and magnon emission/absorption (i.e., longitudinal spin relaxation).

\emph{Anisotropic MR}-In ferromagnets, anisotropic spin relaxation MR ($\ell_{\Vert}\neq \ell_{\perp}$) always 
appears together with the shifted Hanle MR ($\mathcal{B}_{sd}\neq 0$)  even when $B\rightarrow 0$, and thus cannot always be easily separated in experiments. Both  enter through $B$ dependence of  $\Delta
\rho _{1}$ and share the same dependence on the magnetization direction, which adds to the uncertainties of interpreting the experimental data. If Hanle precession is artificially switched off, spin-relaxation anisotropy alone yields  anisotropic MR: when the spin Hall polarization is parallel to the magnetization ($\hat{m}_y=1$), edge spin-$y$ accumulation is preserved (low resistance), while for perpendicular alignment ($\hat{m}_y=0$) it is absorbed (high resistance). Notably, this contribution has higher resistance for lower temperature [see Fig.~\ref{AMR}(a)], distinguishing from the ordinary temperature dependence~\footnote{For conventional resistance, resistance increases with temperature for temperature larger than Kondo temperature.}. For the perpendicular alignment, the edge spin-$y$ accumulation is more easily absorbed by local moments at lower temperature, resulting in higher resistance. Conversely, with isotropic relaxation but finite Hanle precession, anisotropic MR also emerges: a  magnetic field in $\hat{x}$ or $\hat{z}$ direction causes precession of the edge spin-$y$ accumulation, reducing it and thereby increasing resistance.

Quantitatively, the $T$ dependence of $
\Delta\rho _{1}$ is plotted in Fig.~\ref{AMR}(a). The shape of the curve reveals two contributions: spin-exchange field and spin-relaxation anisotropy. The former is reflected in sharp vanishing at $T=T_c$ (blue curve), while the latter is plotted in the inset of Fig.~\ref{AMR}(a) by artificially setting $\mathcal{B}_{sd}=0$.  Both anisotropic and Hanle MR demonstrate themselves by the cosine-square characteristics of $\rho_L$ concerning the magnetization direction $\alpha$ [blue and green curves of Fig.~\ref{AMR}(b)]. At a high  temperature ($T\geq T_c$), no anisotropic MR is observed at $B\rightarrow 0$ [red line of Fig.~\ref{AMR}(b)].  The predicted behaviors qualitatively matches the experimentally observed anisotropic MR~\cite{rushforth2007anisotropic,van2002temperature}. Therefore, we attain a new mechanism of anisotropic MR owing to the interplay of the SHE  and spin relaxation anisotropy.

\begin{figure}[t]
\begin{center}
\includegraphics[width=0.46\textwidth]{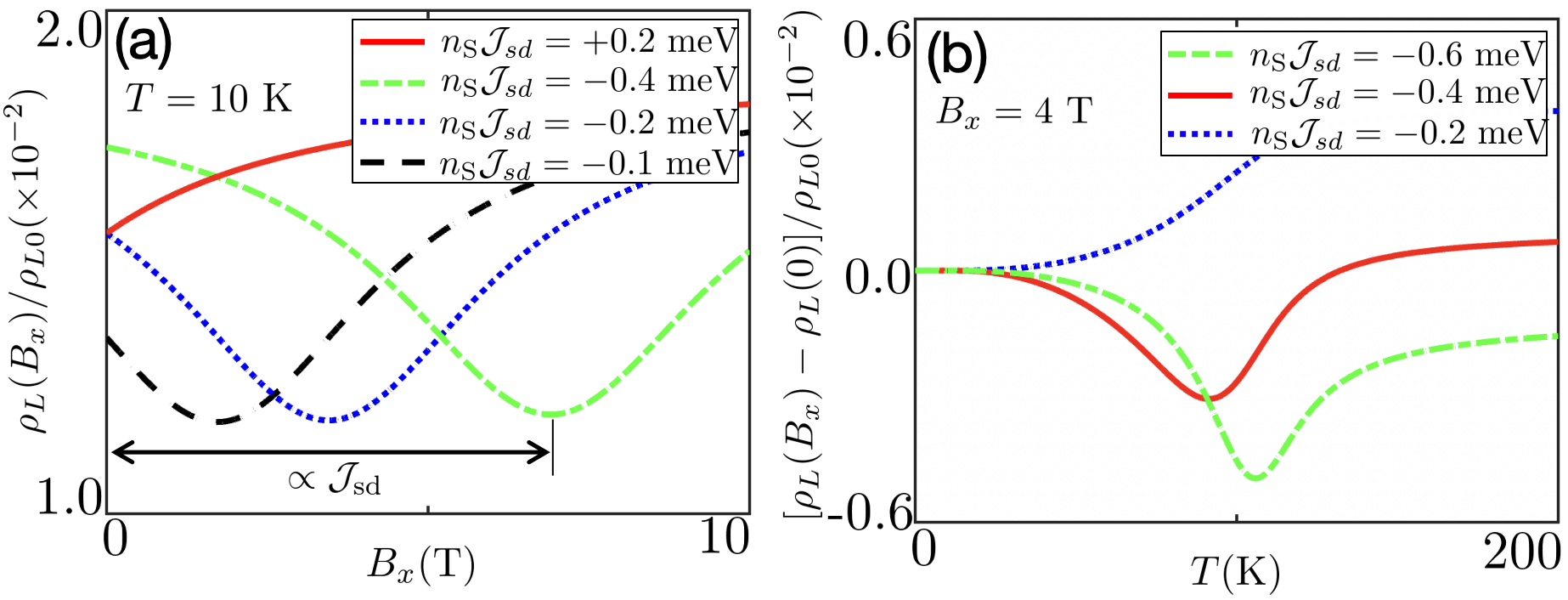} 
\end{center}
\caption{(Color online) Diverse MR behaviors. (a) Resistivity as a function of $\hat{x}$-axis
magnetic field, $B_{x}$, for different values of SEC $\mathcal{J}_{sd}$. (b) $\rho_{\text{L}}$ vs $T$, for various $\mathcal{J}_{sd}$.  The $B$- and $T$-dependent $\mathcal{B}_{sd}$ causes a minimum in resistivity with $B$. Other parameters are the same as FIG.
\protect\ref{AMR}.}
\label{3DVACNJvsH}
\end{figure}

\emph{Diverse behaviors of MR}-Next, we show intriguing MR for the magnetic field in $x$-axis direction ($B_x$). The SEC \eqref{spin-exchange} induces 
a spin-exchange field, which is linearly proportional to the SEC $\mathcal{J}_{sd}$ and the magnetization $
\langle S_{\parallel }\rangle$, i.e., $\mathcal{B}_{sd}\propto \mathcal{J}_{sd}\langle S_{\Vert }\rangle$. Hence, the ferromagnetic (antiferromagnetic) SEC generates a
blue (red) shift of the Hanle spin precession frequency $\omega _{L}=\textsl{g}\mu _{B}(B_x-\mathcal{B}_{sd})/\hbar$ with $\mathcal{B}_{sd}<0$ ($\mathcal{B}_{sd}>0$). At a low enough temperature [black curve in Fig.~\ref{3DVACNJvsH}(a)], i.e., strongly magnetized regime, the spin-exchange field, spin relaxation time \eqref{LongiSRT} and \eqref{TransSRT} acquire their saturated values and become independent of the magnetic field. Then, the magnetic field dependence of resistance purely originates from the magnetic-field spin precession frequency, leading to the shifted Hanle MR $\Delta\rho_{\text{L}}\propto \omega^2 _{L}\propto  (B_x-\mathcal{B}_{sd})^2$ with $
\mathcal{B}_{sd} = -n_{\mathrm{S}}   \mathcal{J}_{sd}   S/(\hbar g\mu _{B})$. The ferromagnetic (antiferromagnetic) SEC leads to positive (negative) MR with $\mathcal{B}_{sd}<0$ $(\mathcal{B}_{sd}>0)$, as indicated by the red and blue curves in Fig.~\ref{3DVACNJvsH}(a). Notably, the minimum of the shifted Hanle MR is where the applied magnetic
field offsets the spin-exchange field, i.e., $B_x=\mathcal{B}_{sd}$.  At saturated temperature ($T=10$ K), the motion of the minimum for different $\mathcal{J}_{sd}$ is plotted in Fig.~\ref{3DVACNJvsH}(a). We observe the spin-exchange field, $\mathcal{B}_{sd}\propto |\mathcal{J}_{sd}|$, is rightly shifted by the stronger antiferromagnetic SEC. This result, in return, can be used to electrically detect the SEC strength, which was previously only detected by the Kondo resonance of scanning tunneling spectroscopy~\citep{wahl2007exchange}. Note that the spin-exchange field also relies on $T$. We find a minimum in resistivity with $T$ for sizable antiferromagnetic SEC like the Kondo effect~\cite{kondo1964resistance}, as shown in Fig.~\ref{3DVACNJvsH}(b).

\emph{Measurement of quantum decoherence}--Finally, we propose an experimental scheme for determining quantum (spin) decoherence ($\ell_{\Vert/\perp}$)—the key factor limiting the functionality of nanoscale quantum devices. As shown in Eqs.~\eqref{vvh0} and \eqref{vvh1}, the decoherence (dephasing and relaxation) is quantified by  $\Delta\rho_0$ and $\Delta\rho_1$, which capture the $\hat{m}$–independent and –dependent components, respectively. The magnon MR measurement yields the $B$ and $T$ dependence of $\Delta\rho_0$, from which we extract  $\ell_{\parallel}$ [c.f.~Eq.~\eqref{vvh0}], which approaches $\ell_0$ in the strongly magnetic regime (e.g., $T \ll T_c$ at $B=0$). Similarly, the anisotropic MR measurement provides the $B$ and $T$ dependence of $\Delta\rho_1$, enabling us to determine $\ell_{\perp}$ and $\ell_{\text{L}}$ [c.f.~Eq.~\eqref{vvh1}]. Consequently, the $B$ and $T$ dependence of $\Delta\rho_0$ and $\Delta\rho_1$—the two key observables in MR experiments—allow direct extraction of quantum decoherence.

\emph{Summary}-We develop a quantum spin decoherence theory of MR, and  microscopically demonstrate diverse MR effects in \textit{mesoscopic} ferromagnets, including magnon, anisotropic, and Hanle MR, which arise from the magnon-induced spin flip, anisotropic spin relaxation, and Hanle spin precession of itinerant electrons, respectively. Our theory provides insights into experimental observations involving $B$- and $T$-dependent MR, paving the way for further progress in the understanding and practical applications of magnetic materials.
Besides, we reveal fruitful behaviors due to the interplay of various MR, allowing electrical detection of the strength of the microscopic SEC. Furthermore, our theory uncovers how MR in mesoscopic ferromagnets serves as an electrical probe of quantum decoherence. Our method can be generalized to more different magnetic materials (e.g.  ferrimagnetic/antiferromagnetic metals/semiconductors), antiferromagnets/altermagnets~\cite{zhang2024electric}, and normal metal decorated with magnetic impurities~\cite{zhang2025open}.

\textit{Acknowledgment --} This work is supported by the National Natural Science Foundation of China (Grant No. 12321004, No. W2511003, No. 12234003, 12505012, No. 12274027, and No. 12374122) and the Hong Kong Research Grants Council Grants (No. 16300523).

\end{document}